\documentclass[12pt]{article}
\usepackage{amsfonts,amsmath,amssymb,mathrsfs}
\title{Some Comments on the Formal Structure of Spontaneous Localization Theories}
\author{Arkadiusz Jadczyk\footnote{
  IMP, Chateau Saint Martin, 82100 Castelsarrasin, France. Email:
  arkadiusz.jadczyk@cict.fr}}
\date{March 6, 2006}

\newcommand{\go}{\mathfrak}
\newcommand{\gH}{\go H}
\newcommand{\RR}{\mathbb{R}}
\newcommand{\PPP}{\mathbb{P}}
\newcommand{\vx} {{\bf x}}
\newcommand{\vX} {{\bf X}}
\newcommand{\va} {{\bf a}}
\newcommand{\gab}{G_{\alpha\beta}}
\newcommand{\gba}{G_{\beta\alpha}}
\newcommand{\ra}{\rho_\alpha}
\newcommand{\ha}{H_\alpha}
\newcommand{\la}{\Lambda_\alpha}
\newcommand{\D}{\mathrm{d}} % differential d
\newcommand{\E}{\mathrm{e}} % exponential e
\newcommand{\I}{\mathrm{i}} % imaginary i
\newcommand{\tr}{\mathrm{Tr}} % imaginary i
\newcommand{\beq}{\begin{equation}}
\newcommand{\eeq}{\end{equation}}
\newcommand{\z}[1]{{#1}}

\begin{document}\maketitle\sloppy

\begin{abstract}
We propose a mathematical and a conceptual framework (called the
``E-model'' and the ``event ontology'' resp.) that encompasses and
generalizes the ``flash'' ontology discussed in a recent paper by
R. Tumulka \cite{tom1}
\medskip

  model.

\end{abstract}

%\maketitle
\section{Introduction}
Tumulka \cite{tom1} proposed a generalization of the GRW model
\cite{GRW} (for a recent overview see \cite{bg1}) and a new
ontology for discussing quantum phenomena, an ontology that he
called ``the flash ontology''. We propose a method of generalizing
Tumulka's idea even further, so as to include a class of models
for which the arena where the ``flashes'' take place may be
different than $3$-space. For instance, our proposed generalized
framework may well include models avoiding explosion of energy as
discussed recently by Bassi et al. \cite{bassi1}, as well as
models described in finite-dimensional Hilbert spaces (e.g. pure
spin models), or models where ``flashes'' are accompanied by phase
transitions. We will also argue that the flash ontology can be
traced back to a more general ``event ontology''. At the end we
will discuss possible advantages of such a generalization and
discuss some applications.

For the purpose of this paper we will call Tumulka's model the
``F-model'' (F - for flash), and we will call our proposed
generalization the ``E-model'' (E - for event). Loosely speaking
``events'' are more general than flashes. While every flash
constitutes an event, we can think of events (for instance
receiving a bit of information) that are not accompanied by
flashes. It should be mentioned though that in \cite{tom1} both
terms, flashes and events, are being used, so our distinction
between the two is rather of an organizing than of a principal
significance.

\section{A brief description of the F-model}
We will describe the model for one particle. Extending it to many
particles, or two quantum field theory, does not represent
essentially new conceptual or technical problems (cf. ``Concluding
Remarks'' at the end of this paper). Let $\gH$ be the Hilbert
space $L^2(\RR^3),$ and let $H$ be a self--adjoint Hamilton
operator in $\gH$ defining the evolution of the particle when the
flash rate is zero. The main object in Tumulka's F-model is a
positive operator valued function $\Lambda(\vx),\, \vx\in\RR^3$.
The physical dimension of $\Lambda(\vx)$ should be
$1/$(time$\times$space). For each Borel set $B\in \RR^3,$
$\Lambda(B)$ denotes the integral of $\Lambda(\vx )$ over $B$, in
particular $\Lambda(\RR^3)$ is the integral of $\Lambda(\vx)$ over
the whole space.\footnote{A similar object has been introduced in
Ref. \cite{ajad1}, but care must be taken when comparing objects
denoted by the same symbols in Refs. \cite{tom1} and \cite{ajad1}.
$\Lambda^{1/2}(\va)$ of \cite{tom1} corresponds to the operator of
multiplication by the function $f_{\va}(\cdot)$ of \cite{ajad1},
while $\Lambda(\RR^3)$ of Ref. \cite{tom1} corresponds to the
operator denoted simply by $\Lambda$ in \cite{ajad1} - cf.
\cite[Eq. (19)]{ajad1}. } The evolution of the wave function
between flashes is given by the evolution operator \beq W_t =
\E^{-\frac{1}{2} \Lambda(\RR^3)t - \I Ht}.\eeq  After each flash,
assuming that the flash happened at the point $\vX$,  the wave
function ``collapses'': \beq \psi \mapsto
\frac{\Lambda(\vX)^{1/2}\psi}{\Vert
\Lambda(\vX)^{1/2}\psi\Vert}.\eeq The joint probability density
distribution of the first $n$ flashes to happen at space--time
points $(T_1,\vX_1),\ldots ,(T_n,\vX_n)$ is postulated to be (cf.
Ref. \cite[Eq. (11)]{tom1}:
\begin{multline}\label{nflashdist}
  \PPP\bigl( \vX_1\in \D^3 \vx_1, T_1 \in \D t_1, \ldots,
  \vX_n \in \D^3 \vx_n, T_n \in \D t_n \bigr) =\\
  \bigl\| K_n(0,\vx_1, t_1, \ldots, \vx_n, t_n) \, \psi \bigr\|^2
  \, \D^3 \vx_1 \, \D t_1 \cdots \D^3 \vx_n \, \D t_n \,,
\end{multline}
where $K_n$ is an operator-valued function defined by
\begin{multline}\label{Kndef}
  K_n(t_0,\vx_1, t_1, \ldots, \vx_n, t_n) =\\
  \Lambda(\vx_n)^{1/2} \,W_{t_n-t_{n-1}} \Lambda(\vx_{n-1})^{1/2}
  \,W_{t_{n-1}-t_{n-2}}
  \cdots \Lambda(\vx_1)^{1/2} \, W_{t_1-t_0} \,.
\end{multline}
The flashes are supposed to be objective space--time events.
Tumulka writes: \begin{quotation}``It is tempting to regard the
collapsed wave function $\psi_t$ as the ontology, but I insist
that the flashes form the ontology.''\end{quotation}
\section{The E--Model}
\subsection{From F-model to E-model}
The E--model can be thought of as a natural and far reaching
extension of the F--model. The concept of a ``flash'' is replaced
by that of an ``event'', with the idea that the model can be
applied to more general systems than systems of particles in
space--time. For instance it may used in quantum cosmology and
applied there to the whole universe. The Big Bang may be such an
``event'', accompanied by some kind of a wave collapse, possibly
with a phase transition. As in the F-ontology, so here events are
thought to be objective, and therefore they can be recorded. It
is, in principle,  possible to analyze the recorded data and to
analyze the past history of the events. It is possible to adjust
the quantum evolution at each given time $t$, depending on the
observed events till the time $t$. Moreover, as we will see, it
costs us nothing to allow for phase transitions that can accompany
some of the events. Formally such a phase transition may involve
change of the Hamiltonian, or even change of the underlying
Hilbert space. Last but not least, we will allow for some sample
histories to have no events at all - this may happen for instance
when an elementary particle passes through a particle detector. It
may well happen that it will cause no detection event at all, due
to the limited detector sensitivity. While the F-model has been
constructed with the purpose of reproducing and generalizing the
GRW spontaneous localization mechanism, the E-model extends the
formal framework of the F-model and enhances it, allowing for a
wider scope of applications, beyond just space--time collapses.
The fact that we allow for dependence of the current evolution on
the past history of events may suggest that we are going beyond
the Markovian models. But this is not the case. It is enough to
formally introduce the recording medium as a part of the system,
and then the evolution becomes Markovian again, while at the same
time adding flexibility and allowing for a wider scope of
applications. We will now introduce the E-model in its full
generality, and later on describe how the F-model becomes a
particular case of the F-model, and how the flash rate density
postulate (\ref{nflashdist}) of Ref. \cite{tom1} can be derived
(rather than postulated) from a simple and natural, Lindblad's
type, Master equation.
\subsection{The formal structure of the E-model}
Let $S$ be a set. Heuristically $S$ is to be thought of as ``the
set of all the potential states of the event recording medium''.
$S$ may have any cardinality required, but for the sake of
simplicity we will assume in this Section that $S$ is  a finite
set. Generalization to more complicated case (such as, for
instance, the GRW model, or the F-model of Ref. \cite{tom1}) is
straightforward: it requires replacing sums by integrals etc.
Elements of $S$ will be denoted by Greek letters $\alpha,\beta ,$
etc. For every $\alpha\in S$ let there be given a Hilbert space
$\gH_\alpha .$ In most of the applications (when there are no
phase transitions) all Hilbert space $\gH_\alpha$ can be
identified: $\gH_\alpha \equiv \gH,\, \forall \alpha\in S.$ For
each $\alpha\in S$ let $H_\alpha (t)$ be a self-adjoint
Hamiltonian operator acting in $\gH_\alpha .$ For the sake of
generality we allow $H_\alpha$ to depend explicitly on time $t\in
\RR.$ We define an ``event'' to be an ordered pair
$(\alpha,\beta),\, \alpha\neq\beta,\, \alpha,\beta\in S.$ Thus,
heuristically, an event is defined as a ``change of state of the
event recording medium'' -- a natural definition. The final piece
of data consists in associating to each event
$\alpha\rightarrow\beta$ a transition operator
$G_{\beta\alpha}:\gH_\alpha\longrightarrow\gH_\beta.$ For
convenience we will extend the definition of $\gab$ to the case of
``non-events'', $\beta=\alpha$ by putting  $G_{\alpha\alpha}\doteq
0,\, \forall\alpha .$ In Ref.\cite{tom1} an event is a flash at
$\vX\in\RR^3,$ and the associated jump (or ``collapse'') operator
is $\Lambda(\vX)^{\frac{1}{2}}.$ Here the transition operators are
the primitive objects, while the positive operators $\Lambda$ are
composite. The relation between $G$-s and $\Lambda$-s is
many--to--one. It is important to notice that it is, in general,
not necessary for these transition (or ``jump'')  operators to be
positive, and we will allow for this more general case in our
model. As with the Hamiltonians $H_\alpha,$ so with the jump
operators, we can allow $\gba(t)$ to depend explicitly on time.
Assuming the initial state is described by a vector
$\psi_\alpha\in \gH_\alpha,$ the sample history will be piecewise
deterministic: there will be a continuous evolution
$\psi_\alpha(t)$ between jumps, interrupted with a sequence of
transitions $\alpha_k\rightarrow\alpha_{k+1}$ at discrete times
$T_k,$ accompanied by quantum jumps
$\gH_{\alpha_k}\ni\psi_k(T_k)\rightarrow\psi_{k+1}(T_k)\in\gH_{\alpha_{k+1}}.$
Averaging over these discrete, random jumps, the initial pure
state $\psi_\alpha$ will evolve continuously into a mixture state
with components in different $\gH_{\alpha_k}.$ Let
$\rho=\{\rho_\alpha\},\, \rho_\alpha\geq 0\, \forall\alpha ,\, \tr
(\rho )=\sum_\alpha \tr (\rho_\alpha)=1,$ be the density matrix of
our system. Then we postulate the following evolution equation as
the basic equation for the E--model: \beq
\frac{\D\,\rho_\alpha}{\D\, t}\, =-i[\ha,\ra]+\sum_\beta \gab\,
\rho_\beta\, \gab^\star - \frac{1}{2}\{\la,\ra\},\label{eq:liour}
\eeq where \beq \la=\sum_\beta \gba^\star \gba, \eeq and
$G_\alpha,H_\alpha$ may depend explicitly on $t.$ By taking the
trace and summing over $\alpha$ we find that $\tr (\rho )$ is
conserved. Moreover, Eq. (\ref{eq:liour}) is of Lindblad's type,
thus positivity is preserved as well. While Eq. (\ref{eq:liour})
describes the average behavior of a statistical ensemble, a
typical sample history is described by a simple piecewise
deterministic Markov process described in what follows. Let us
introduce the
following notation:\\
For each $\alpha\in S$ and each pair $t_0<t$ of time instants, let
$W_\alpha(t,t_0)$ be  the unique solution of the differential
equation:\footnote{We disregard possible complications coming from
 the necessity of taking care of the domains of definition of the involved generators} \beq {\dot W_\alpha(t,t_0)}=\left( -\I H_\alpha
(t)-\frac{1}{2}\Lambda_\alpha(t)\right)
W_\alpha(t,t_0),\label{eq:p}\eeq with the initial condition
$W_\alpha(t,t_0){\vert_{t=t_0}}=I_\alpha,$
 the identity operator on $\gH_\alpha.$ For each $t\in\RR,$ $\beta\neq \alpha,$
$\psi\in\gH_\alpha,$ let \beq p_{\alpha\rightarrow\beta}(\psi
,t)=\frac{\langle\psi|G_{\beta\alpha}(t)^\star
G_{\beta\alpha}(t)\psi \rangle}{\langle\psi , \Lambda_\alpha (t)
\psi\rangle} =\frac{\Vert \gba(t)\psi\Vert^2}{\sum_\gamma \Vert
G_{\gamma\alpha}(t)\psi\Vert^2} .\label{eq:pab} \eeq The sample
path of the unique piecewise deterministic Markov process
reproducing Eq. (\ref{eq:liour}) is then described as
follows:\begin{quotation} Given on input $t_0,\alpha_0,$ and
$\psi_0\in{\go H}_{\alpha_0}$, with $\Vert \psi_0\Vert=1$, it
produces on output $t_1,\alpha_1$ and $\psi_1\in{\cal
H}_{\alpha_1}$, with $\Vert\psi_1\Vert=1$. \vskip 0.2cm
\noindent $1$) Choose uniform random number $r\in [0,1]$.\\
\noindent $2$) Propagate $\psi_0$ in ${\go H}_{\alpha_0}$ forward
in time: \beq \psi(t)= W_\alpha(t,t_0)\psi_0\eeq  until $t=t_1$,
where $t_1$ is defined by\footnote{Note that, as can be seen from
the equation (\ref{eq:p}), due to non--negativity of the operators
$\Lambda_\alpha(t),$ the norm of $\psi(t)$ is a monotonically
decreasing function of $t$.} \beq \Vert \psi(t_1) \Vert^2=r.
\label{eq:time} \eeq \noindent $3$) Select the new state
$\alpha_1\in S,$ among the states $\alpha\neq\alpha_0,$ using the
probability
distribution $p_{{\alpha_0}\rightarrow\alpha}\left(\psi(t_1),t_1\right).$\\
\noindent $4$) Set \beq \psi_1=G_{\alpha_1\alpha_0}(t_1)\psi(t_1)/
\Vert G_{\alpha_1\alpha_0}(t_1)\psi(t_1) \Vert \in
\gH_{\alpha_1}\eeq. Goto $1)$ replacing $\alpha_0$ with
$\alpha_1,$ $t_0$ with $t_1$ and $\psi_0$ with
$\psi_1.$\end{quotation}
 \vskip 0.2cm \noindent The sample history of an individual
system is described by a repeated application of the above
algorithm, using its output as the input for each next step.

As a Corollary we easily get the following joint distribution of
the first $n$ events:
\begin{multline}\label{neventsdist}
  \PPP\bigl( \alpha_1, T_1 \in \D t_1, \ldots,
  \alpha_n, T_n \in \D t_n \bigr) =\\
  \bigl\| K_n(\alpha_0,t_0,\alpha_1, t_1, \ldots, \alpha_n, t_n) \, \psi_0 \bigr\|^2
  \, \D t_1 \cdots \, \D t_n \,,
\end{multline}
where $K_n$ is an operator-valued function on
$(S\times\mathrm{time})^{n+1}, $
$K_n:\gH_{\alpha_0}\rightarrow\gH_{\alpha_n},$ defined by
\begin{multline}\label{Knmydef}
  K_n(\alpha_0,t_0,\alpha_1, t_1, \ldots, \alpha_n, t_n) =\\
  G_{\alpha_n\alpha_{n-1}}(t_n) \,W_{\alpha_{n-1}}(t_n,t_{n-1})
  G_{\alpha_{n-1}\alpha_{n-2}}(t_{n-1})
  \,W_{\alpha_{n-2}}(t_{n-1},t_{n-2})
  \cdots\, G_{\alpha_1\alpha_0}(t_1)W_{\alpha_0}(t_1,t_0)\,
  .
\end{multline}
The derivation of Eq. (\ref{neventsdist}) from the Markov process
is straightforward. According to step $1)$ the process resulting
in the first jump is an inhomogeneous Poisson process with rate
function \beq \lambda (t)=\frac{\langle
W_{\alpha_0}(t,t_0)\psi_0\vert\Lambda_{\alpha_0}(t)W_{\alpha_0}(t,t_0)\psi_0\rangle}{\langle
W_{\alpha_0}(t,t_0)\psi_0\vert W_{\alpha_0}(t,t_0)\psi_0\rangle }
.\eeq The probability of surviving without any event until $t_1$
is $\Vert W_{\alpha_0}(t_1,t_0)\Vert^2,$ and the probability of an
event during the time interval $t_1$ and $t_1+\D t$ is $\lambda
(t_1)\D t$, so that the probability that the first jump will occur
between $t_1$ and $t_1+\D t_1$ is $$\Vert
W_{\alpha_0}(t_1,t_0)\Vert^2\times\lambda (t_1)\D t =\langle
W_{\alpha_0}(t_1,t_0)\psi_0\vert\Lambda_{\alpha_0}(t_1)W_{\alpha_0}(t_1,t_0)\psi_0\rangle\D
t.$$ Now, the probability of a transition
$\alpha_0\rightarrow\alpha_1$ at $t_1$ is
$$p_{{\alpha_0}\rightarrow{\alpha_1}}(t_1,W_{\alpha_0}(t_1,t_0)\psi_0)=\frac{\Vert
G_{\alpha_1\alpha_0}(t_1)W_{\alpha_0}(t_1,t_0)\psi_0\Vert^2}{\langle
W_{\alpha_0}(t_1,t_0)\psi_0\vert\Lambda_{\alpha_0}(t_1)W_{\alpha_0}(t_1,t_0)\psi_0\rangle
},$$ and so the probability that the first event will happen
between $t_1$ and $t_1+\D t,$ and will be accompanied by the
transition $\alpha_0\rightarrow\alpha_1$ is the product
\begin{multline}\langle
W_{\alpha_0}(t_1,t_0)\psi_0\vert\Lambda_{\alpha_0}(t_1)W_{\alpha_0}(t_1,t_0)\psi_0\rangle\D
t\times \frac{\Vert
G_{\alpha_1\alpha_0}(t_1)W_{\alpha_0}(t_1,t_0)\psi_0\Vert^2}{\langle
W_{\alpha_0}(t_1,t_0)\psi_0\vert\Lambda_{\alpha_0}(t_1)W_{\alpha_0}(t_1,t_0)\psi_0\rangle}=\\
=\Vert
G_{\alpha_1\alpha_0}(t_1)W_{\alpha_0}(t_1,t_0)\psi_0\Vert^2\D t.
\end{multline}
Repeating the steps, starting now with
$$\psi_1=\frac{G_{\alpha_1,\alpha_0}(t_1)W_{\alpha_0}(t_1,t_0)\psi_0}{\Vert
G_{\alpha_1,\alpha_0}(t_1)W_{\alpha_0}(t_1,t_0)\psi_0\Vert }$$
leads to equations (\ref{neventsdist}) and (\ref{Knmydef}).
\section{From E--model to F--model}
A sequence of specializations leads from the E--model described
above to the F--model of Ref. \cite{tom1}:\begin{enumerate}
\item The set $S$ is specialized to be the set of all finite
sequences $\{\vx_1,\ldots ,\vx_n\},\, \vx_i\in\RR^3.$ If $\alpha$
and $\beta$ are two such sequences, then $G_{\beta\alpha}\neq 0$
only if $\beta$ can be obtained from $\alpha$ by adding one space
point at the end: $\alpha=\{\vx_1,\ldots ,\vx_n\}$ and
$\beta=\{\vx_1,\ldots ,\vx_n,\vx_{n+1}\}.$ An event
$\alpha\rightarrow\beta$ is then interpreted as a ``flash'' at
$\vx_{n+1}.$
\item Assume that all Hilbert spaces are identical
$\gH_\alpha=\gH,$ that there is only one, time--independent
Hamiltonian $H_\alpha(t)=H,$ and that the transition operators do
not depend neither on the initial point $\alpha\in S,$ nor on time
$t$: $G_{\alpha,\alpha_0}(t)=G_\alpha.$ Notice that if this is the
case, then $\Lambda_\alpha(t)=\sum_\beta G_\beta^\star
G_\beta\equiv \Lambda$ does not depend neither on the index
$\alpha$ nor on time $t.$ The operators $W_\alpha (t,t_0)$ do not
depend now on $\alpha,$ they depend only on the difference $t-t_0$
and are given by $W_\alpha (t,t_0)=W_(t-t_0),$ where
$$W_t =\E^{\left( -\I
H-\frac{1}{2}\Lambda\right)t}$$. Notice that, with these
assumptions, introducing $\rho=\sum_\alpha \rho_\alpha,$ the
Master equation (\ref{eq:liour}) can be now summed over $\alpha$
and leads to the evolution equation for $\rho$: \beq
\frac{\D\,\rho}{\D\, t}\, =-i[H,\rho ]+\sum_\beta G_\beta\, \rho,
G_\beta^\star - \frac{1}{2}\{\Lambda,\rho\},\label{eq:liouref}.
\eeq
\item With $\alpha=\{\vx_1,\ldots ,\vx_n\}$ let us furthermore
assume that $G_\alpha$ depends only on the last element of the
sequence $G_{\{vx_1,\ldots ,\vx_n\}}=G_{\vx_n}.$ Assume, moreover,
that the operators $G_{\vx}$ are non--negative.
\end{enumerate}
With the above specializations, identifying $G_{\vx}^2$ of the
E--model with $\Lambda(\vx )$ of the F-model, and replacing the
sum in Eq. (\ref{neventsdist}) by the integral, we recover the
joint probability distribution, and thus the flash generating
process of Ref. \cite{tom1}.\footnote{In the F--model the index
${\bf a}$ of $G_{\bf a}$ is now continuous, and we can modify the
integral over ${\bf a}$ on a set of measure zero, without changing
the process.}
\section{Concluding Remarks}
Tumulka begins his paper \cite{tom1} discussing the flash ontology
with the following statement: \begin{quotation} John S.~Bell
concluded from the quantum measurement problem that ``either the
wave function, as given by the Schr\"odinger equation, is not
everything or it is not right'' \cite{Belljumps}. Let us assume,
for the purpose of this paper, the second option of the
alternative [...]\end{quotation} But, in fact, the flash ontology,
and to even greater extent the event ontology discussed above
require the first option as well: the wave function is not
everything. Flashes, and more generally and more distinctively,
events form a separate ontology.
\subsection{Possibility of a chaotic behavior}
In general the operators $\Lambda(\vx )$ of Ref. \cite{tom1} do
not have to form a commutative family. They commute for a GRW
model, where they are all functions of the standard
quantum--mechanical position operator of the particle. But, for
instance,  Bassi et al. \cite{bassi1} suggested models in which
the transition operators can be functions of the position {\emph
and}\, momentum operators. Cases with non--commuting transition
operators have been studied numerically within a simple class of
E--models - cf. Ref \cite{ajad2}, where it was shown that the
resulting patterns of events may show a chaotic, fractal--like
behavior.
\subsection{Taming the energy increase} While I was demonstrating a
computer simulation of the flash generating process during the
recent conference in honor of GianCarlo Ghirardi's 70th birthday,
Roderich Tumulka pointed to me that the average time distance
between subsequent flashes seems to be getting smaller and
smaller. I did not notice it before, so after the conference has
ended I wrote a specialized computer program to investigate just
this phenomenon. Lo and behold, Tumulka was right. It is only then
that I have discovered that the phenomenon is known to the
experts, and that the recent paper of Bassi at al. \cite{bassi1}
emerged from the discussion of this phenomenon, and from the
attempt to provide a cure for the possible infinite increase of
the velocity of the particle as measured by the frequency of
flashes. The cure seems to consist of mixing space and momentum
localizations. Such a generalization is available within the
E--model. Replacing space by phase space, and replacing
multiplication operators by Gaussian functions of $\vx$, as in GRW
model, with Wigner quantized Gaussian functions of the position
$\vx$ and the momentum ${\bf p}$ may lead to a model along the
lines indicated in \cite{bassi1}. Wigner's quantization does not
preserve positivity, but positivity of the transition operators is
not required in the E--model. The fact that the resulting
operators will then form a non--commuting family may lead to an
extra chaotic behavior of the observed trajectories.
\subsection{Quantum Field Theory\label{sec:qft}} Generalization of the proposed
framework to the case of many particles (whether distinguishable
or not), and then further to quantum field theory, is
straightforward as it has been shown in Refs. \cite{ajad1} and
\cite{tom1}.

\bigskip

\noindent\textit{Acknowledgments.} \z{I would like to thank the
organizers of the conferences ``Are there quantum Jumps'' and ``On
the present status of quantum mechanics'', the Consorzio per la
Fisica, and the QFG for a partial support of this work.}


\begin{thebibliography}{9}

\bibitem{tom1}
Roderich Tumulka, \emph{On Spontaneous Wave Function Collapse and
Quantum Field Theory}, Preprint - {\emph arXiv:quant-ph/0508230}.

\bibitem{GRW} Ghirardi, G. C., Rimini, A., Weber, T.: ``Unified
  dynamics for microscopic and macroscopic systems'', Phys. Rev. D
  \textbf{34}, 470--491 (1986)

\bibitem{bg1} A.~Bassi, G.~C.~Ghirardi: ``Dynamical
  Reduction Models'', {\emph Phys. Rep.} \textbf{379}, (2003) 257--427  and
  {\emph arXiv:quant-ph/0302164}

\bibitem{bassi1}
A.~Bassi, E.~Ippoliti, B.~Vacchini, ``On the Energy Increase in
Space--Collapse Models'', \emph{J. Phys. A: Math. Gen. }
\textbf{38}, (20005) 8017--8038 , and {\emph arXiv:
quant-ph/0506083}.

\bibitem{ajad1} A.~Jadczyk, ``On Quantum Jumps, Events and
Spontaneous Localization Models'', {\emph Found. Phys.}
\textbf{25}, (1995), 743--762 , and {\emph arXiv:hep-th/9408021}

\bibitem{ajad2} A. ~Jadczyk, ``Simultaneous Measurement of
Non--commuting observables and Quantum Fractals on Complex
Projective Spaces'', {\emph. Chinese J. of Physics} {\textbf 43}
no {\textbf 2}, (2005) 301--327, and {\emph
arXiv:quant-ph/0311081}

\bibitem{Belljumps} J.~S.Bell: ``Are there quantum jumps?'', in
  \textit{Schr\"odinger. Centenary of a polymath}, p.~41--52.
  Cambridge: Cambridge University Press (1987). Reprinted in
  \cite{Bellbook}, p.~201--212.

\bibitem{Bellbook} J.~S. Bell: \textit{Speakable and unspeakable in
  quantum mechanics}.  Cambridge: Cambridge University Press (1987)

\end{thebibliography}
\end{document}